\documentclass[runningheads]{llncs}
\usepackage{amssymb}
\usepackage{subcaption, multirow, array}
\usepackage{booktabs}
\usepackage{xspace}
\usepackage{mathtools}
\usepackage{siunitx}
\usepackage{orcidlink}
\usepackage[misc]{ifsym}

\mathchardef\mhyphen="2D
\newcolumntype{C}{@{\extracolsep{6pt}}c@{\extracolsep{3pt}}}%
\newcolumntype{L}{@{\extracolsep{6pt}}l@{\extracolsep{3pt}}}%
\newcommand{\specialcell}[2][l]{%
  \begin{tabular}[#1]{@{}l@{}}#2\end{tabular}}
\newcommand{\method}{AngioMoCo}
  
\makeatletter
\DeclareRobustCommand\onedot{\futurelet\@let@token\@onedot}
\def\@onedot{\ifx\@let@token.\else.\null\fi\xspace}

\makeatother
\newcommand{\subpara}[1]{\vspace{3pt} \noindent \textbf{#1.}}

\usepackage[T1]{fontenc}
\usepackage{graphicx}
\usepackage{hyperref}
\usepackage{color}

\begin{document}
\title{\method: Learning-based Motion Correction \\in Cerebral Digital Subtraction Angiography}

%
\author{
Ruisheng Su\inst{1,6} \thanks{r.su@erasmusmc.nl} \orcidlink{0000-0002-5013-1370}\and
Matthijs van der Sluijs\inst{1} \orcidlink{0000-0002-4934-0933} \and
Sandra Cornelissen\inst{1} \orcidlink{0000-0002-0332-2158}\and
Wim van Zwam\inst{2} \orcidlink{0000-0003-1631-7056} \and
Aad van der Lugt\inst{1} \orcidlink{0000-0002-6159-2228} \and
Wiro Niessen\inst{1,3} \orcidlink{0000-0002-5822-1995}\and
Danny Ruijters\inst{4} \orcidlink{0000-0002-9931-4047} \and
Theo van Walsum\inst{1} \orcidlink{0000-0001-8257-7759} \and
Adrian Dalca\inst{5,6} \orcidlink{0000-0002-8422-0136}
}

\authorrunning{Su et al.}

\institute{
Erasmus University Medical Center, The Netherlands \and 
Maastricht University Medical Center, The Netherlands \and
Delft University of Technology, The Netherlands \and
Philips Healthcare, The Netherlands \and
Massachusetts Institute of Technology, Boston, USA \and
Massachusetts General Hospital, Harvard Medical School, USA
}

\titlerunning{AngioMoCo: Learning-based Motion Correction in DSA}

\maketitle     
\begin{abstract}
Cerebral X-ray digital subtraction angiography (DSA) is the standard imaging technique for visualizing blood flow and guiding endovascular treatments. The quality of DSA is often negatively impacted by body motion during acquisition, leading to decreased diagnostic value. Time-consuming iterative methods address motion correction based on non-rigid registration, and employ sparse key points and non-rigidity penalties to limit vessel distortion. Recent methods alleviate subtraction artifacts by predicting the subtracted frame from the corresponding unsubtracted frame, but do not explicitly compensate for motion-induced misalignment between frames. This hinders the serial evaluation of blood flow, and often causes undesired vasculature and contrast flow alterations, leading to impeded usability in clinical practice. To address these limitations, we present \method, a learning-based framework that generates motion-compensated DSA sequences from X-ray angiography. \method{} integrates contrast extraction and motion correction, enabling differentiation between patient motion and intensity changes caused by contrast flow. This strategy improves registration quality while being substantially faster than iterative elastix-based methods. We demonstrate \method{} on a large national multi-center dataset (MR CLEAN Registry) of clinically acquired angiographic images through comprehensive qualitative and quantitative analyses. \method{} produces high-quality motion-compensated DSA, removing motion artifacts while preserving contrast flow. Code is publicly available at \url{https://github.com/RuishengSu/AngioMoCo}.\par

\keywords{Angiography \and X-Rays \and Registration \and Motion Artifacts.}
\end{abstract}
\section{Introduction}
\begin{figure}[!t]
\centering
\includegraphics[clip, trim=0cm 0cm 0cm 0cm, width=\textwidth]{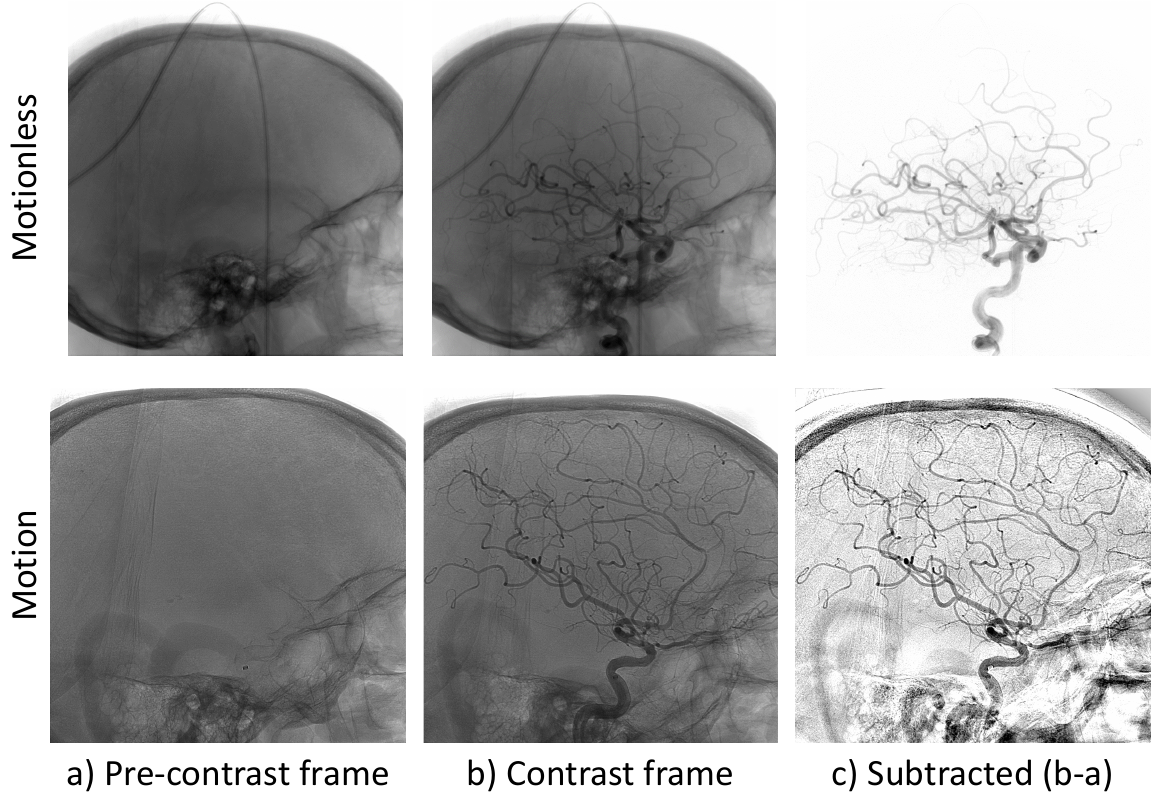}
\caption{Illustration of motion artifacts in DSA: a) the pre-contrast frame; b) a subsequent post-contrast frame; c) subtracted frame (b-a).} \label{fig:dsa_examples}
\end{figure}

Cerebral X-ray digital subtraction angiography (DSA) is a widely used imaging modality in interventional radiology for blood flow visualization and therapeutic guidance in endovascular treatments~\cite{shaban2021digital}. It is a 2D+T image series obtained by subtracting an initial pre-contrast image from subsequent post-contrast frames, leaving only the contrast-filled vessels visible. The injection of contrast medium and the subtraction process effectively eliminate soft tissue and bone, enabling high-resolution visualization of the vessels and the blood flow. However, this subtraction technique assumes the absence of motion between frames during exposure. In clinical practice, this premise is often violated. Involuntary motions, caused by swallowing, coughing, stroke, or endovascular procedures, are nearly inevitable. Body motion results in undesired artifacts in subtracted images, leading to decreased image quality and impaired interpretability of DSA (Fig.~\ref{fig:dsa_examples}).\par 


Over the last three decades, various motion correction techniques have been proposed to mitigate the impact of body motion retrospectively~\cite{meijering1999retrospective}. Registration algorithms typically employ template matching with corresponding control points or landmarks to align images~\cite{bentoutou20053,bentoutou2002invariant,buzug1998image,buzug2006using,cao2005dsa,chu2006registration,cox1994automatic,liu2013stretching,meijering2001reduction,meijering1999image,nejati2013nonrigid,staring2007rigidity,sundarapandian2013dsa,taleb1998image}. These algorithms rely on features based on vessels~\cite{cao2005dsa}, edges~\cite{chu2006registration,meijering2001reduction,meijering1999image,taleb1998image}, corners~\cite{wang2009iterative}, textures~\cite{nejati2010fast}, temporal correspondence~\cite{bentoutou20053}, and non-uniform grids~\cite{sundarapandian2013dsa}. Some methods capture both local and global transformations, such as multi-resolution search~\cite{nejati2014multiresolution,yang2007multiresolution}, block matching~\cite{chu2006registration}, and iterative estimations~\cite{nejati2010fast,wang2009iterative}. Others limit undesirable vessel distortions, such as sparse key points~\cite{meijering1999image} and non-rigidity penalties~\cite{staring2007rigidity}. Although these methods are effective in motion compensation, they require time-consuming iterative computation for each frame, limiting their clinical applicability.\par

Recent generative learning-based models, such as pix2pix~\cite{isola2017image}, have been adapted to address subtraction artifacts without registration~\cite{crabb2022deep,gao2019deep,ueda2021deep}. These models leverage deep learning techniques to predict a subtraction image from an input post-contrast image by discerning foreground contrast from the body background, resulting in reduced artifacts. However, these models do not explicitly compensate for motion-induced misalignment between frames, often cause hallucinations or modification of contrast and vessels, and lack interpretability. Consequently, these shortcomings hinder the serial evaluation of blood flow and impede the diagnostic utility of DSA.\par

To overcome these limitations, we introduce \method{}, a straightforward, fast, and effective learning-based motion correction method for DSA that avoids severe contrast distortion. We employ a supervised CNN module that as a preliminary step distinguishes between motion displacement and contrast intensity change. The output contrast-removed image and the pre-contrast image are used as input to a subsequent self-supervised learning-based registration model for deformable registration, where a deformation regularization loss limits the local irregularity. By excluding contrast enhancements from the deformation learning process, \method{} avoids undesired distortion of the vessels. The resulting warp is used to correct the original post-contrast image. This results in trustworthy visualization of continuous blood flow and promises to assist in automated analysis of flow-based biomarkers relevant to endovascular treatments.\par

Overall, classical non-rigid registration methods use various regularization strategies to limit vessel distortion, but are prohibitively time-consuming. Recent learning-based methods are fast, but do not explicitly model the motion between frames, and as a result can negatively distort or hallucinate the clinical information we aim to highlight. We build on the strengths of both directions while avoiding their limitations. Specifically, we propose a novel learning-based strategy that is significantly faster than traditional non-rigid registration methods. \method{} not only removes subtraction artifacts on each frame but does so by explicitly compensating for motion between frames, which is not available in existing image-to-image models. We demonstrate that  \method{} achieves high-quality registration while avoiding undesirable contrast reduction or vessel erasure.\par


\section{Method}

\subsection{Model}
\begin{figure}[!t]
\centering
\includegraphics[clip, trim=0cm 0cm 0cm 0cm, width=\textwidth]{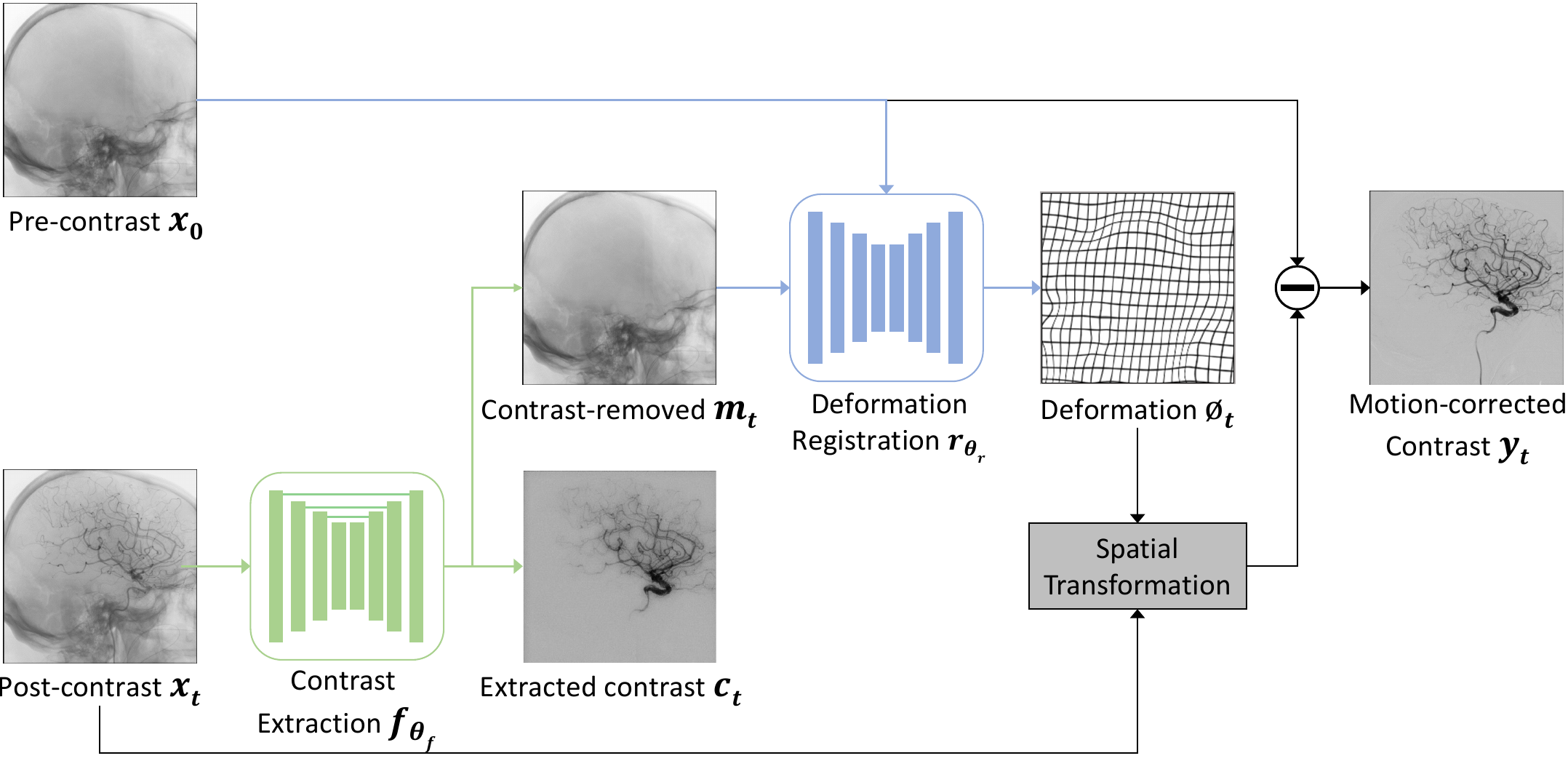}
\caption{\textbf{Overview.} \method{} takes a pre-contrast image $x_0$ and a post-contrast image $x_t$ as input. The contrast extraction module $f_{\theta_f}(\cdot)$ splits $x_t$ into contrast $c_t$ and contrast-removed $m_t = x_t - c_t$. Next, $m_t$ and $x_0$ are registered using network $r_{\theta_r}(\cdot, \cdot)$, which outputs a deformation field $\phi_t$. Subsequently, $\phi_t$ is applied to the post-contrast image $x_t$ to obtain the final output subtracted image $y_t$, which corrects misalignment between frames.} \label{fig:architecture}
\end{figure}

Fig.~\ref{fig:architecture} outlines the \method{} framework for motion correction and subtraction in angiographic images, comprising three main modules: initial contrast extraction, deformable registration, and spatial-transformed subtraction. Let $\mathcal{X} = \{x_t\}_{t=0}^{T}$ be the 2D+T DSA series of a patient, where $x_0$ is the pre-contrast frame and $\{x_t\}_{t=1}^{T}$ are the post-contrast frames.\par

We define a contrast extraction module $f_{\theta_f}(x_t) = c_t$ with parameters $\theta_f$ that takes as input a post-contrast frame $x_t$. This function separates $x_t$ into a contrast image $c_t$ and a contrast-removed image $m_t$ where $m_t = x_t - c_t$. The values in $c_t$ are within [-1, 0] as the injected contrast medium can only lead to a decrease in pixel intensity relative to the input image with an intensity range of [0, 1]. The contrast extraction module aims to reduce contrast discrepancies between the pre- and post-contrast frames. \par

Such image-to-image modules can lead to hallucination and may not fully capture distal vessels, relatively less contrasted vessels, and vessels behind bone structures. Therefore, in \method{}, we only employ this module to enable easier registration of the frame $x_t$ to the pre-contrast $x_0$ using the intermediate contrast-extracted $m_t$ image as a proxy.\par 

We define a registration function $r_{\theta_r}(x_0, m_t) = \phi_t$ with parameters $\theta_r$ to estimate the deformation field $\phi_t$. We then obtain the motionless subtraction angiography $y_t$ by subtracting the pre-contrast frame $x_0$ from the warped post-contrast frame $w_t$: 
\begin{align}
{y_t} &= w_t - x_0\\
      &= x_t \circ \phi_t - x_0 ,
\label{eq:y_t}
\end{align}
where $\circ$ defines a spatial warp.\par

\subsection{Training}

We train the contrast extraction $f_{\theta_f}(\cdot)$ and deformable registration $r_{\theta_r}(\cdot, \cdot)$ modules separately. We train the contrast extraction module on a motionless subset of the train data with an MSE loss between the ground truth contrast, estimated via subtraction between post- and pre-contrast frames ($x_t - x_0$), and the predicted $c_t$:
\begin{align}
\mathcal{L}_{\text{ext}}(\theta_r; x_t) &= \mathcal{L}_{\text{MSE}}(x_t-x_0, f_{\theta_f}(x_t)) .
\label{eq:L_modular_ext}
\end{align}
We train the deformable registration module on a motion subset of the train data, with the pre-trained contrast extraction module frozen, using a loss function that combines an MSE loss between $m_t$ and $x_0$ and a smoothness loss $\mathcal{L}_{\text{smooth}}$, weighted by $\lambda$:
\begin{align}
\mathcal{L}_{\text{reg}}(\theta_f; x_0, m_t\circ \phi_t) &= (1-\lambda)\mathcal{L}_{\text{MSE}}(x_0, m_t\circ \phi_t) + \lambda \mathcal{L}_{\text{smooth}}(\phi_t) ,
\label{eq:L_modular_reg}
\end{align}
where $\mathcal{L}_{\text{smooth}}$ is the mean squared horizontal and vertical gradients of displacement $u_t$ in deformation field $\phi_t$, that enforces the deformation spatial smoothness:
\begin{align}
\mathcal{L}_{\text{smooth}}(\phi_t) &= \|\nabla u_t\|^2.
\end{align}

\subsection{Architecture}
We design the contrast extraction module $f_{\theta_f}(\cdot, \cdot)$ using a U-Net architecture, which includes a contracting path (encoder) and an expanding path (decoder) connected by skip connections. The encoder stage comprises eight convolutional and max-pooling layers with the number of channels being 8, 16, 32, 64, 128, 256, 512, and 512 respectively. The convolutions operate with a 3x3 kernel size and a stride of 2. Similarly, the decoding path employs eight upsampling, 3x3 convolution, and concatenation operations with 32 feature maps per layer to restore the spatial dimension up to the input size. Each convolution is accompanied by an instance normalization and a LeakyReLU activation layer. We also use three additional 3x3 convolutions. The final convolution employs a negative sigmoid activation, confining the output pixel intensity to [-1, 0].\par

We employ a deformable registration module $r_{\theta_r}(\cdot,\cdot)$ based on VoxelMorph to learn motion correction in DSA~\cite{balakrishnan2019voxelmorph}. We add instance normalization between the convolution layers of the encoder and decoder. We use this deformable registration module to predict bi-directional dense deformation fields using diffeomorphism that easily enables to spatial transformation of either pre- or post-contrast frames.\par




\section{Experiments}
We assess \method{} in terms of vessel contrast preservation, artifact removal, and computation efficiency compared to existing approaches.\par

\subsection{Experimental Setup}
\subsubsection{Data.}
We identified 272 patients with unsubtracted cerebral angiographic images available from MR CLEAN registry~\cite{jansen2018endovascular}, an ongoing prospective observational multi-center registry of patients with acute ischemic stroke who underwent endovascular thrombectomy (EVT). This comprised 788 angiographic series, consisting of 16,641 frames in total, acquired between attempts of thrombus retrieval. The DSA series were acquired using various imaging systems, including Philips, GE, and Siemens, and had a size of $1024\times1024$ pixels. The series had varying lengths, ranging from 10 to 50 frames, and temporal resolutions between 0.5 and 4 frames per second (fps). We performed image resizing to 512$\times$512 pixels and min-max intensity normalization to obtain intensity values within the range of $[0, 1]$. To ensure the coherency of the intensity along the series during normalization, the maximum intensity is calculated on the series level based on the stored bits in the DICOM header.\par

Based on visual assessment, we categorized the dataset into two subsets: motionless and motion. We use the motionless subset, consisting of 107 series (1933 frames) from 21 patients, for pre-training and evaluating the contrast extraction module. The motion subset, which contains 681 series (14708 frames) from 251 patients, is used for overall training and evaluation. We split data on the patient level independently on the motionless and motion subsets, with a ratio of 50\%, 20\%, and 30\% for training, validation, and testing, respectively.\par

\subpara{Baselines.}
We compare \method{} with two widely used image registration approaches, elastix-based affine registration and VoxelMorph~\cite{balakrishnan2018unsupervised,balakrishnan2019voxelmorph}, and an image-to-image approach employing a U-Net~\cite{ronneberger2015u} architecture. We followed the implementation of~\cite{balakrishnan2019voxelmorph} for VoxelMorph. For the U-Net, we employed the same architecture as the contrast extraction module $f_{\theta_f}(\cdot, \cdot)$ with the same preprocessing and augmentations. We trained the U-Net using the motionless subset and used mean squared error (MSE) as the optimizing objective. We implemented the methods using Python 3.10.6 and PyTorch~\cite{paszke2019pytorch}.\par

\subpara{Training details.} 
We use an NVIDIA 2080 Ti GPU (11 GB), the Adam optimizer~\cite{kingma2014adam} and the ReduceLROnPlateau scheduler with an initial learning rate of 0.001, a patience of 300 epochs, and a decay of 0.1. We set the batch size to 8 and applied early stopping with a patience of 500 epochs. We selected these optimization parameters based on validation performance using a grid search. We show results for several deformation regularization $\lambda$. We applied data augmentations using Albumentations~\cite{buslaev2020albumentations}, including \textit{HorizontalFlip}, \textit{ShiftScaleRotate}, and \textit{RandomSizedCrop}, each with a probability of 0.5. 

\subpara{Evaluation.}
We carry out both qualitative and quantitative analyses on the hold-out test set of the motion subset. A key challenge is to minimize motion and subtraction artifacts while retaining clinically important features. We use mean squared intensity (MSI) as a proxy to quantify the preservation of contrast intensity within vessels and the ability of motion correction outside vessels. As ground truth deformations are not available for image sequences with motion, we manually segment the blood vessels in post-contrast frames (Supplemental Fig. 6), and use the resulting masks to quantify MSI inside and outside blood vessels. We used paired t-tests for statistical significance.

\subsection{Results}
\begin{figure}[!t]
\centering
\includegraphics[clip, trim=0cm 0cm 0cm 0cm, width=\textwidth]{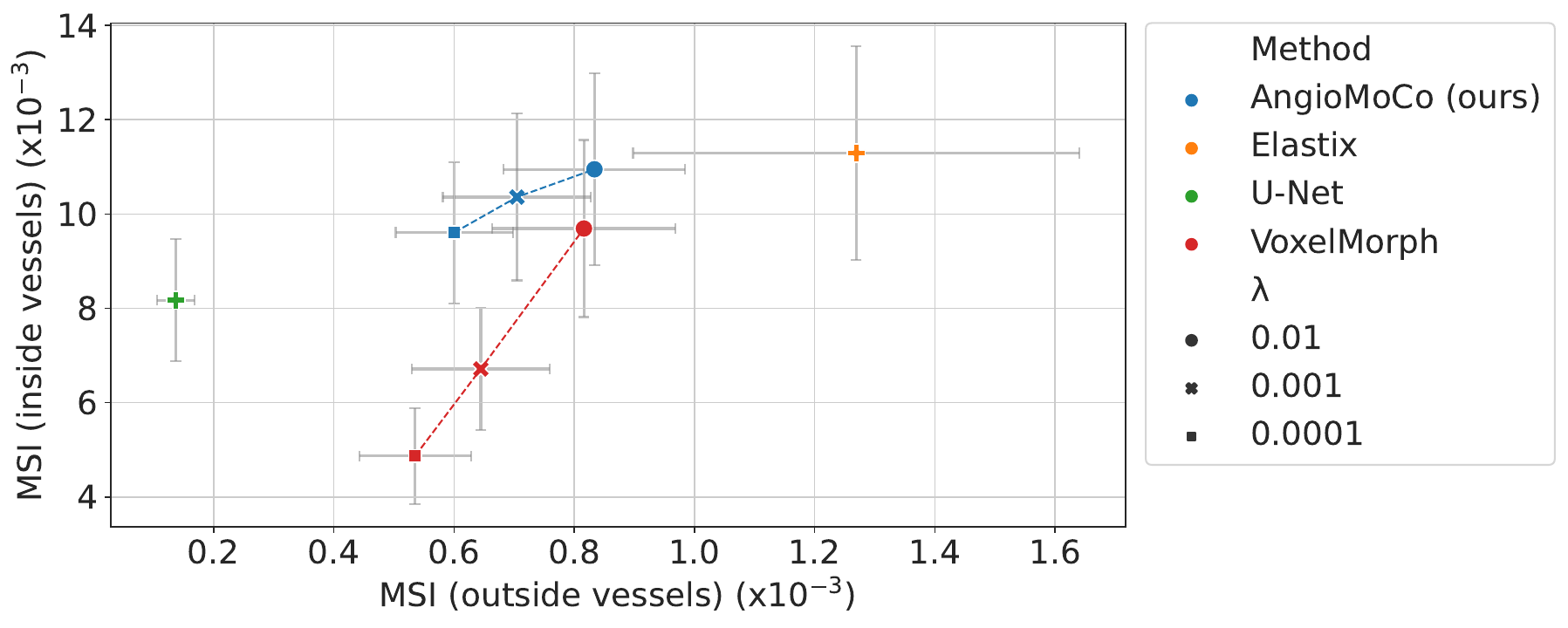}
\caption{Mean squared intensity (MSI) on the test set. Better methods will preserve the MSI (i.e., vessel contrast) inside vessels ($\uparrow$, y-axis) while minimizing the MSI (i.e., artifacts) outside vessels ($\leftarrow$, x-axis), moving towards the top left of the graph.} \label{fig:quantitative_results}
\end{figure}

\subpara{Quantitative analysis}
%
The optimal outcome is represented by the top left corner of Fig.~\ref{fig:quantitative_results}, indicating high vessel contrast preservation and complete artifact removal (Supplemental Table 1). We illustrate several results of \method{}, corresponding to different values of the core registration hyperparameter $\lambda$. Compared to elastix affine registration, \method ($\lambda=0.001$) achieves similar vessel preservation (P=0.2), while substantially decreasing the MSI outside vessels (by about half). Compared to VoxelMorph, \method{} demonstrates substantial improvement, with higher vessel preservation and better (more to the left) artifact removal. While the image-to-image U-Net yields the lowest MSI outside vessels, it sacrifices a substantial amount (30\%) of contrast inside vessels, harming the precise clinical signal we are interested in.\par

\begin{figure}[!t]
\centering
\includegraphics[clip, trim=0cm 0cm 0cm 0cm, width=\textwidth]{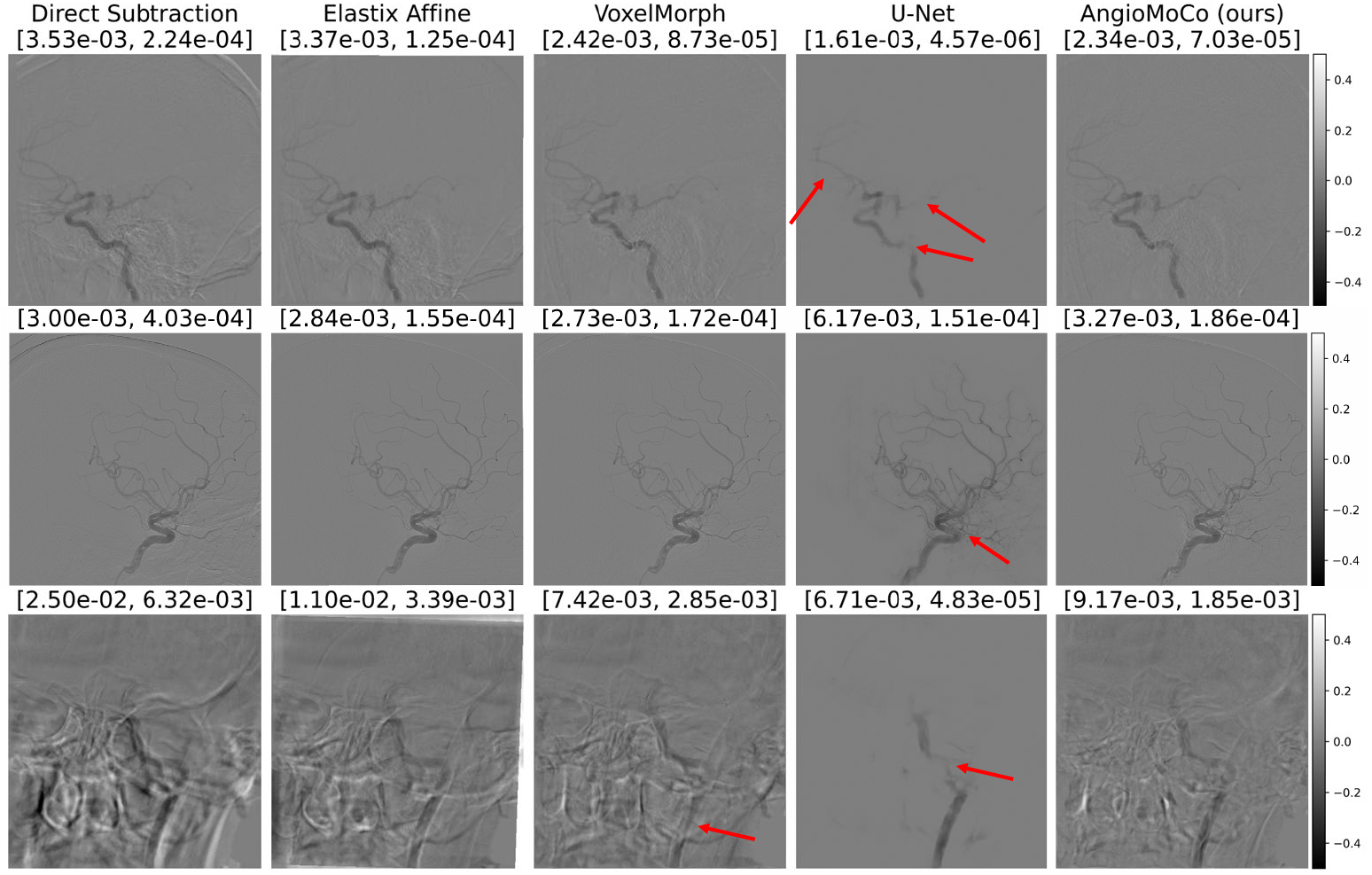}
\caption{Representative visual comparisons. We report MSI values inside (left) and outside (right) vessels in brackets. Red arrows point to undesired vessel contrast erasure or modifications. \method{} achieves better background artifact removal and vessel enhancement than other methods. The UNet achieves excellent artifact removal, but it comes at the cost of severe damage to the vessels of interest, making it clinically less useful.} \label{fig:qualitative_comparison}
\end{figure}

\subpara{Qualitative analysis}
Figure~\ref{fig:qualitative_comparison} presents visual comparisons of the methods through three representative examples. The image-to-image U-Net generates images with fewer motion artifacts than other methods, but it often fails to capture vessel contrast behind bone structures (Row 1), distal vessels (Row 1), and loses high-frequency spatial features, leading to blurry images (Row 2). These errors can have substantial negative effects on downstream clinical applications. VoxelMorph operates on pre- and post-contrast images, which can cause considerable modifications in the vessel contrast flow. For example, the motion-corrected image of VoxelMorph in Row 3 has lighter vessel contrast than its counterparts. In contrast, \method{} overcomes these limitations of U-Net and VoxelMorph by learning to disentangle contrast flow from motion.\par 

\subpara{Runtime}
Compared to existing iterative registration methods running on the CPU only, deep-learning-based registration methods, including \method{}, require orders of magnitude less time. For example, \method{} takes less than a second to process a series on GPU, while iterative registration methods are mostly implemented on CPU where they require minutes.\par


\section{Discussion}

We find that \method{} achieves high-quality motion correction in DSA, while preserving vessel details, which is of critical clinical importance. While the image-to-image U-Net resulted in fewer artifacts, it substantially degrades the vessel contrast, harming its usability in clinical usefulness.\par

These results suggest that \method{} is clinically relevant for endovascular applications, enhancing the utility of DSA in diagnosis and treatment planning. The tool can extract contrast flow while outputting smooth bi-directional deformation fields that provide interpretability. Unlike image-to-image models, the contrast flow visualization is driven by motion-compensation of the post-contrast frames to the pre-contrast image, and hence avoids undesirable hallucinations and modifications of vessel contrast. 

We also examined the end-to-end training strategy of \method, which did not yield superior results to VoxelMorph or the modularly trained \method{} (Supplemental Fig. 5). To further enhance registration accuracy, future research may explore the integration of 3D spatio-temporal CNN and the utilization of vessel masks as auxiliary supervision.\par



\section{Conclusion}
We presented \method, a deep learning-based strategy towards motion-free digital subtraction angiography. The approach leverages a contrast extraction module to disentangle contrast flow from body motion and a deformable registration module to concentrate on motion-induced deformations. The experimental results on a large clinical dataset demonstrate initial findings that \method{} outperforms iterative affine registration, learning-based VoxelMorph, and image-to-image U-Net. Overall, \method{} achieves high registration accuracy while preserving vascular features, improving the quality and clinical utility of DSA for diagnosis and treatment planning in endovascular procedures.\par

\section*{Acknowledgments}
This work is supported by Health-Holland (TKI Life Sciences and Health) through the Q-Maestro project under Grant EMCLSH19006 and Philips Healthcare (Best, The Netherlands). This work is done during a visit at MGH. The visit was made possible in part by the Academy Van Leersum grant of the Academy Medical Sciences Fund of the Royal Netherlands Academy of Arts \& Sciences (KNAW). The work was also supported by NIH grants R01AG064027 and R01AG070988. \par

%
%
\bibliographystyle{splncs04}
\bibliography{paper522}

\newpage 
\section*{Appendix A} \label{sec:appendix_a}
\begin{table}[!ht]
\centering
\caption{Overview of mean squared intensity (MSI) on 50 test patients. We aim to preserve the MSI (i.e., vessel contrast) inside vessels ($\uparrow$) while minimizing the MSI (i.e., artifacts) caused by motion outside vessels ($\downarrow$). We show the results of \method{} with regularization $\lambda=0.001$. The table displays the mean MSI in the first row and the corresponding 95th percentile range in the second row.}\label{tab:quantitative_results}
\begin{tabular}{lLLLLL}
\hline
Region & \specialcell{Direct subtraction\\$\times10^{-3}$} & \specialcell{Elastix (affine)\\$\times10^{-3}$} & \specialcell{VoxelMorph\\$\times10^{-3}$} & \specialcell{U-Net\\$\times10^{-3}$} & \specialcell{\method\\$\times10^{-3}$} \\ [0.2cm] \hline

\multirow{2}{*}{\specialcell{Inside\\vessels}\hspace{0.25cm}$\uparrow$} & 12.37 & 11.30 & 9.69 & 8.17 & 10.36 \\  

& {\scriptsize [7.80, 16.94]} & {\scriptsize [6.85, 15.75]} & {\scriptsize [6.01, 13.37]} & {\scriptsize [5.63, 10.71]} & {\scriptsize [6.89, 13.83]} \\ [0.2cm]  

\multirow{2}{*}{\specialcell{Outside\\vessels}\hspace{0.05cm}$\downarrow$} & 1.79 & 1.27 & 0.82 & 0.14 & 0.70 \\  

& {\scriptsize [0.86, 2.72]} & {\scriptsize [0.52, 1.11]} & {\scriptsize [0.54, 2.00]} & {\scriptsize [0.08, 0.20]} &  {\scriptsize [0.46, 0.95]} \\ [0.2cm] 
\hline

\multicolumn{6}{l}{\footnotesize{\parbox{12cm}{Inside vessels: AngioMoCo vs affine P = 0.2, VoxelMorph vs affine P = 0.002.}}} \\

\multicolumn{6}{l}{\footnotesize{\parbox{12cm}{Outside vessels: AngioMoCo vs others P < 0.05.}}}

\end{tabular}
\end{table}

\section*{Appendix B} \label{sec:appendix_b}
\begin{figure}[!h]
\centering
\includegraphics[clip, trim=0cm 0cm 0cm 0cm, width=\textwidth]{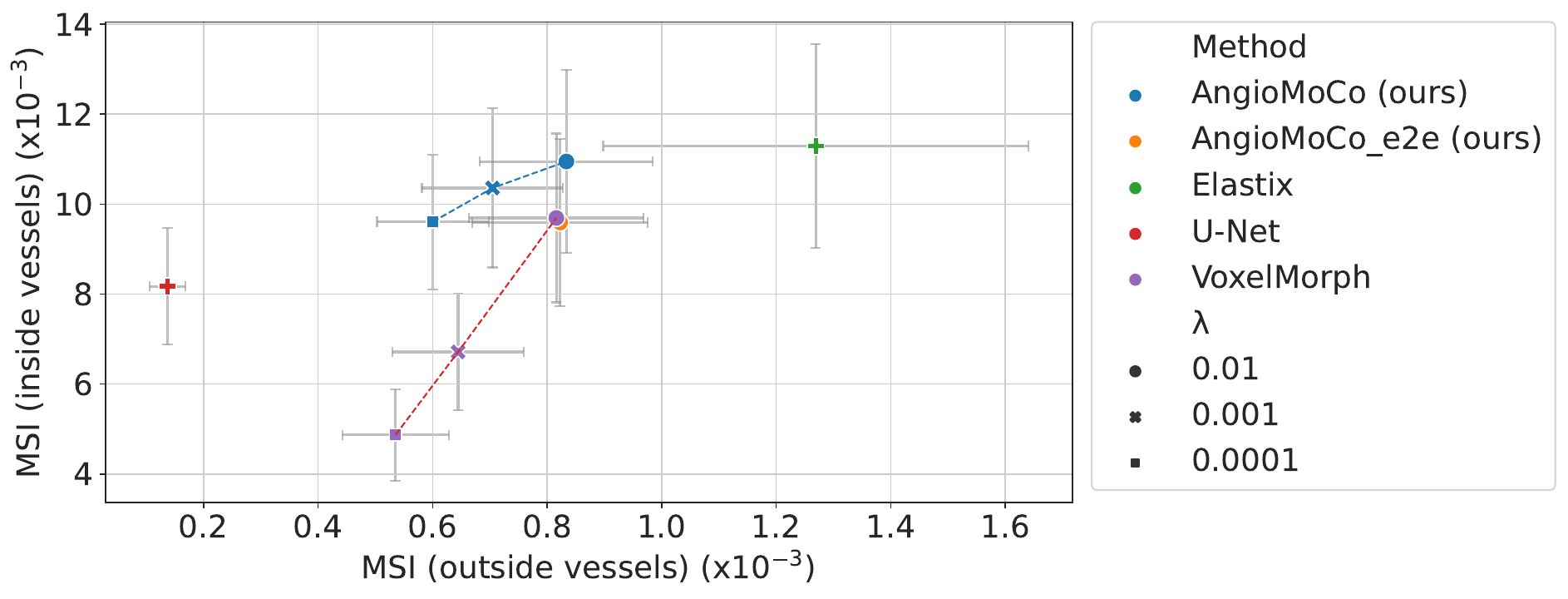}
\caption{Mean squared intensity (MSI) on the test set. Better methods will preserve the MSI (i.e., vessel contrast) inside vessels ($\uparrow$, y-axis) while minimizing the MSI (i.e., artifacts) caused by motion outside vessels ($\leftarrow$, x-axis). \method\_e2e is trained end-to-end on the motion subset of data, which achieves similar performance to VoxelMorph.} \label{fig:quantitative_results_appendix}
\end{figure}

\newpage
\section*{Appendix C} \label{sec:appendix_c}
\begin{figure}[!h]
\centering
\includegraphics[clip, trim=0cm 0cm 0cm 0cm, width=\textwidth]{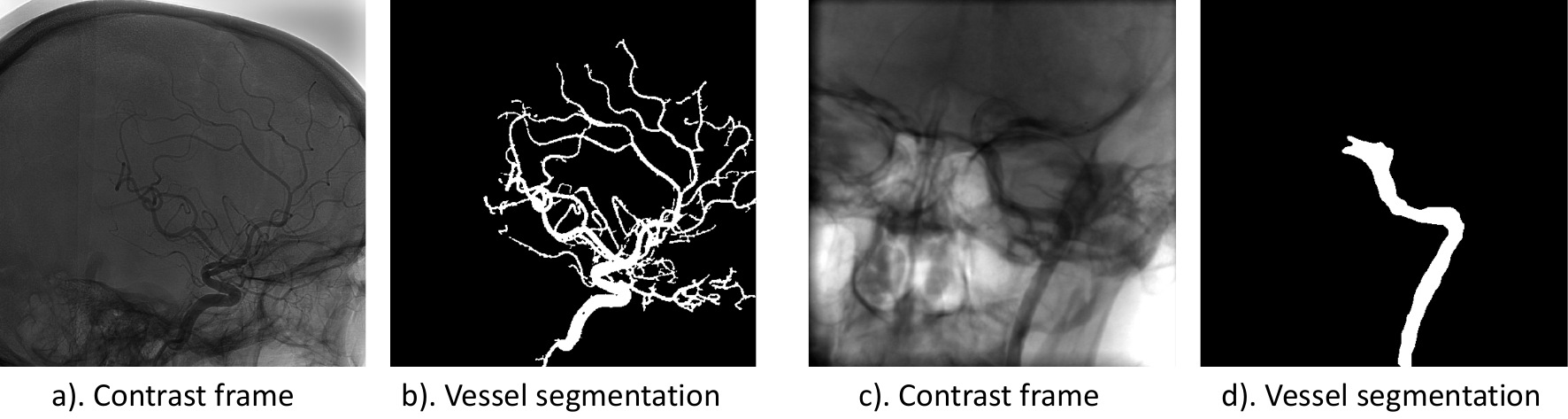}
\caption{Vessel segmentation examples.} \label{fig:segmentation_examples_appendix}
\end{figure}

\end{document}